\documentclass[twocolumn,showpacs,preprintnumbers]{revtex4}%
\usepackage{amsfonts}
\usepackage{amsmath}
\usepackage{graphicx}
\usepackage{dcolumn}
\usepackage{bm}
\usepackage{amssymb}%
\begin{document}
\title{Role of the tetraquark in the chiral phase transition}
\author{Achim Heinz$^{\text{(a)}}$, Stefan Str\"{u}ber$^{\text{(a)}}$, Francesco
Giacosa$^{\text{(a)}}$and Dirk H. Rischke$^{\text{(a,b)}}$}
\affiliation{$^{\text{(a)}}$Institute for Theoretical Physics, Goethe University,
Max-von-Laue-Str.\ 1, D--60438 Frankfurt am Main, Germany }
\affiliation{$^{\text{(b)}}$Frankfurt Institute for Advanced Studies, Goethe University,
Ruth-Moufang-Str.\ 1, D--60438 Frankfurt am Main, Germany }

\begin{abstract}
We investigate the implications of a
light tetraquark field on chiral symmetry restoration at nonzero temperature
within a simple chirally symmetric model.
In order for the chiral phase transition to be crossover, as shown by lattice
QCD studies, a strong mixing between scalar quarkonium and tetraquark fields
is required. This leads to a light ($\sim0.4$ GeV), predominantly tetraquark
state, and a heavy ($\sim1.2$ GeV), predominantly quarkonium state in the
vacuum, in accordance with recently advocated interpretations of spectroscopy
data. The mixing even increases with temperature and leads to an interchange
of the roles of the originally heavy, predominantly quarkonium state and the
originally light, predominantly tetraquark state. Then, as expected, the
scalar quarkonium is a light state when becoming degenerate in mass with the
pion as chiral symmetry is restored at nonzero temperature.

\end{abstract}

\maketitle



\bigskip

\emph{Introduction--} In the last 30 years theoretical and experimental work
on the light scalar mesons with mass below $\sim1.8$ GeV initiated an intense
debate about their nature. The issue is that too many scalar resonances have
been identified than can be accommodated in a naive quark-antiquark picture.
For instance, in the scalar isoscalar channel there are five states:
$f_{0}(600)$, $f_{0}(980)$, $f_{0}(1370)$, $f_{0}(1500)$, and $f_{0}(1710)$
\cite{pdg}. In order to explain their nature, quarkonia, tetraquark and
mesonic molecular assignments, as well as a scalar glueball state with mass
around 1.5 GeV suggested by lattice studies of quantum chromodynamics (QCD)
\cite{lattglue}, have been investigated in a variety of
combinations and mixing patterns \cite{amslerrev,fariborz1,fariborz,tqmix}.
Yet, a clear answer to the question which resonance should be dominantly
identified as a scalar quarkonium (i.e., quark-antiquark) state is not at hand.

Nowadays evidence for a full nonet of scalars with mass below 1 GeV is
mounting; these are the already mentioned isoscalars $f_{0}(600)$ and
$f_{0}(980)$, as well as the isotriplet $a_{0}(980)$, and the two isodoublets
of $K_{0}^{*}(800)$. As proposed long ago by Jaffe \cite{jaffe}, a tetraquark
assignment for these states can explain some puzzling properties, such as the
mass ordering which is reversed compared to the expectation from a
quark-antiquark picture, and the strong coupling of $a_{0}(980)$ and
$f_{0}(980)$ to kaons \cite{maiani}. Within this context the lightest scalar
resonance $f_{0}(600)$ is interpreted as a tetraquark state $\frac{1}%
{2}[u,d][\overline{u},\overline{d}]$ where the commutator indicates an
antisymmetric flavor (as well as color) configuration of the diquark. Further
indications of a non-quarkonium nature of the scalar states below 1 GeV, and
thus of $f_{0}(600)$, are obtained from a large-$N_{c}$ study in the framework
of unitarized chiral perturbation theory \cite{pelaez} and in the lattice
studies of Ref.\ \cite{mathur}.

If the light scalars are (predominantly) tetraquark states, the question is
how to identify the lightest quarkonium state $\overline{n}n=1/\sqrt{2}
(\overline{u}u+\overline{d}d)$: the broad resonance $f_{0}(1370)$ is the first
candidate. This assignment is also supported by the fact that the scalar
quarkonia are $p$-wave states, and thus expected to lie above 1 GeV together
with other $p$-wave quarkonia such as axial-vector and tensor mesons. The two
isoscalars of the quarkonia nonet and the scalar glueball can mix, forming the
states $f_{0}(1370),$ $f_{0}(1500),$ and $f_{0}(1710)$; such scenarios have
been discussed in Refs.\ \cite{refs}. While it is not clear if $f_{0}(1500)$
or $f_{0}(1710)$ carries the largest glueball amount, all the above cited
works agree in the assignment of a dominant $\overline{n}n$ component to the
resonance $f_{0}(1370)$.

For $N_{f}$ massless quark flavors, the QCD Lagrangian [including effects from
the $U(1)_{A}$ anomaly \cite{tHooft}] has a chiral $SU(N_{f})_{r}\times
SU(N_{f})_{l}\times U(1)_{V}$ symmetry, $V=r+l$. If this symmetry is linearly
realized, the mass eigenstates of QCD come in degenerate pairs, so-called
chiral partners, which have the same quantum numbers except for parity and
G-parity. For instance, the chiral partners of the scalar isoscalar meson are
the pseudoscalar isotriplet mesons, i.e., the pions. The chiral symmetry is
spontaneously broken to $SU(N_{f})_{V}$ in the vacuum \cite{vafawitten}, thus
lifting the degeneracy of the chiral partners and rendering the pions
Goldstone bosons. Since the pion is commonly regarded to be a quark-antiquark
state, and if the resonance $f_{0}(1370)$ is predominantly a quarkonium state,
the latter should be considered as the chiral partner of the pion, and not
$f_{0}(600)$, as is usually assumed. Consequently, if the chiral symmetry of
QCD is restored above some critical temperature $T_{c}$, as predicted by
lattice QCD calculations \cite{karschfodor}, the resonance $f_{0}(1370)$, and
not $f_{0}(600)$, should become degenerate in mass with the pion. Clarifying
this issue is important for the interpretation of data from
heavy-ion-collision experiments, whose major goal is to identify signatures
for chiral symmetry restoration at nonzero temperature $T$.

The aim of this paper is a first step towards investigating this scenario of
chiral symmetry restoration. We employ the toy model discussed in
Ref.\ \cite{tqmix}, which is the $N_{f}=2$ limit of a more general chiral
Lagrangian for $N_{f}=3$. This model contains only one tetraquark field in
addition to the scalar quarkonium field and the pions. Although the other
mentioned scalar-isoscalar resonances $f_{0}(980),$ $f_{0}(1500),$ and
$f_{0}(1710)$ are not included at the present stage, this model has all
the essential features to analyze the role of the tetraquark and its mixing
with the quarkonium at nonzero temperature. To this end we employ the
Cornwall-Jackiw-Tomboulis (CJT) formalism \cite{cjt} in the Hartree-Fock
approximation \cite{juergen}. We shall show that, as $T$ increases, the
$f_{0}(1370$) becomes lighter and its tetraquark admixture grows. Within our
model calculation, we also find that, for a large range of parameters, there
exists a certain temperature $T_{s}\leq T_{c}$ above which the state that is
predominantly quarkonium becomes lighter than the state that is mostly
tetraquark. At and above $T_{c}$, the state which is predominantly quarkonium
becomes degenerate with the pion, as expected for chiral symmetry restoration.

\emph{The model-- }We consider the pion triplet $\vec{\pi},$ the bare
quarkonium field $\varphi\equiv\overline{n}n$ and the bare tetraquark field
$\chi\equiv\frac{1}{2}[u,d][\overline{u},\overline{d}]$. The potential
defining our model emerges as the $SU(2)_{r}\times SU(2)_{l}$ limit of a more
general $SU(3)_{r}\times SU(3)_{l}$ chiral invariant Lagrangian studied in
Ref.\ \cite{tqmix} and reads explicitly:
\begin{equation}
V=\frac{\lambda}{4}(\varphi^{2}+\vec{\pi}^{2}-F^{2})^{2}-\varepsilon
\varphi+\frac{1}{2}M_{\chi}^{2}\chi^{2}-g\chi\left(  \varphi^{2}+\vec{\pi}%
^{2}\right)  \;, \label{vpot}%
\end{equation}
where $\varepsilon$ parametrizes explicit chiral symmetry breaking by nonzero
quark masses and $g$, chosen to be $\geq0$, is the interaction strength of the
tetraquark field $\chi$ [which is a singlet under $SU(2)_{r}\times SU(2)_{l}$]
with the quarkonia fields. As we shall see, $g$ also determines the mixing of
the scalar fields. When $g \rightarrow0$ a simple linear sigma model for
$\varphi$ and $\vec{\pi}$ is left. In fact, the field $\chi,$ with mass
$M_{\chi}$, decouples in this limit.
The minimum of the potential (\ref{vpot}) is, to order $O(\epsilon)$, assumed
for
\begin{equation}
\varphi_{0}\simeq\frac{F}{\sqrt{1-2g^{2}/(\lambda M_{\chi}^{2})}}%
+\frac{\varepsilon}{2\lambda F^{2}}\,,\;\;\chi_{0}=\frac{g}{M_{\chi}^{2}%
}\varphi_{0}^{2}\,, \label{cond}%
\end{equation}
and $\vec{\pi}=0$. The $\bar{n}n$ condensate $\varphi_{0}$ is identified with
the pion decay constant $f_{\pi}=92.4$ MeV. Note that the vacuum expectation
value ($vev$) $\chi_{0}$ is proportional to $\varphi_{0}^{2}$. Thus, the
tetraquark condensate $\chi_{0}$ is induced by the spontaneous symmetry
breaking in the quarkonium sector. After shifting the fields $\varphi
\rightarrow\varphi_{0}+\varphi$ and $\chi\rightarrow\chi_{0}+\chi$ and
expanding the potential around the minimum, we obtain up to second order in
the fields
\begin{equation}
V=\frac{1}{2}(\chi,\varphi)\left(
\begin{array}
[c]{cc}%
M_{\chi}^{2} & -2g\varphi_{0}\\
-2g\varphi_{0} & M_{\varphi}^{2}%
\end{array}
\right)  \left(
\begin{array}
[c]{c}%
\chi\\
\varphi
\end{array}
\right)  +\frac{1}{2}M_{\pi}^{2}\vec{\pi}^{2}+\ldots\,, \label{vexp}%
\end{equation}
where $M_{\varphi}^{2}=\varphi_{0}^{2}\left(  3\lambda-\frac{2g^{2}}{M_{\chi}^{2}%
}\right)  -\lambda F^{2},\text{ }M_{\pi}^{2}=\frac{\varepsilon}{\varphi_{0}}$.
The value $M_{\pi}=0.139$ GeV is used. Due to the off-diagonal terms in the
mass matrix in Eq.\ (\ref{vexp}), the fields $\varphi$ and $\chi$ are not mass
eigenstates of the potential $V$. The latter, denoted by $(H,S)$, are obtained
after an $SO(2)$ rotation of the fields $(\varphi,\chi)$
\begin{equation}
\left(
\begin{array}
[c]{c}%
H\\
S
\end{array}
\right)  =\left(
\begin{array}
[c]{cc}%
\cos\theta_{0} & \sin\theta_{0}\\
-\sin\theta_{0} & \cos\theta_{0}%
\end{array}
\right)  \left(
\begin{array}
[c]{c}%
\chi\\
\varphi
\end{array}
\right)  \,, \label{diag2}%
\end{equation}
where $\theta_{0}=\frac{1}{2}\arctan\frac{4g\varphi_{0}}{M_{\varphi}^{2}-M_{\chi}%
^{2}}\,$.
The tree-level masses of $H$ and $S$ are:%
\begin{align*}
M_{H}^{2}  &  =M_{\chi}^{2}\cos^{2}\theta_{0}+M_{\varphi}^{2}\sin^{2}%
\theta_{0}-2g\varphi_{0}\sin(2\theta_{0}),\\
M_{S}^{2}  &  =M_{\varphi}^{2}\cos^{2}\theta_{0}+M_{\chi}^{2}\sin^{2}%
\theta_{0}+2g\varphi_{0}\sin(2\theta_{0}).
\end{align*}
Assuming $-\pi/4\leq\theta_{0}\leq\pi/4$, the state $H$ is then predominantly
tetraquark and $S$ predominantly quarkonium. As discussed above, we shall
identify the state $H$ with the resonance $f_{0}(600)$ and the state $S$ with
$f_{0}(1370)$. A natural choice is then that the pure tetraquark should be
lighter than the pure quarkonium, i.e., $M_{\chi}<M_{\varphi}$. Using the fact
that the trace and the determinant of the mass matrices before and after the
$SO(2)$ rotation are equal, we obtain
$\left(  M_{S}^{2}-M_{H}^{2}\right)  ^{2}=\left(  M_{\varphi}^{2}-M_{\chi}%
^{2}\right)  ^{2}+\left(  4g\varphi_{0}\right)  ^{2}\,$, implying that for $g>0$ the masses of the states $H$ and $S$ repel each other:
$M_{H}<M_{\chi}<M_{\varphi}<M_{S}$ and that $\left\vert
M_{S}^{2}-M_{H}^{2}\right\vert \geq4g\varphi_{0}$.

Note that we refrain from a more elaborate study of other vacuum properties of
the model (such as decay widths), since a realistic description of the latter
requires the inclusion of other scalar states and of (axial-)vector
mesons \cite{kr,stefan,thermal}. 

As a side remark, we mention that
the tree-level decay width of $f_0(600)$ is larger than 300 MeV when
the mass lies above 0.6 GeV and when the mixing is large. 
However, we refrain from a more elaborate study of vacuum properties, since a realistic description of the latter
requires the inclusion of other scalar states and of (axial-)vector
mesons \cite{kr,stefan,thermal} and we
concentrate on the behavior at 
nonzero $T$.

\emph{Results-- } In order to study chiral symmetry restoration at nonzero $T$
we employ the CJT-formalism in the Hartree-Fock approximation; for details 
see Ref.\ \cite{juergen}. As a result, the masses $M_{H}(T),$ $M_{S}(T)$,
$M_{\pi}(T)$, and the mixing angle $\theta(T)$ become functions of $T$.
Moreover, both scalar-isoscalar fields attain $T$-dependent vacuum expectation
values (vev's), $\varphi_{0}\rightarrow\varphi(T)$ for the quarkonium and $\chi
_{0}\rightarrow\chi(T)$ for the tetraquark state, respectively, with
$\varphi(0)=\varphi_{0}$ and $\chi(0)=\chi_{0}$, see Eq.\ (\ref{cond}).

We first study the order of the chiral phase transition and the associated
$T_{c}$ as a function of the model parameters $g$, $M_{H} \equiv M_{H}(0)$,
and $M_{S} \equiv M_{S}(0)$. Figure 1(a) shows the phase diagram in the
$g$-$M_{S}$ plane for fixed $M_{H}=0.4$ GeV which is close to the value of
Ref.\ \cite{caprini}, while Fig.\ 1(b) depicts the $g$-$M_{H}$ plane for fixed
$M_{S}=1.2$ GeV which is in the experimentally established range of values for
$f_{0}(1370)$ \cite{pdg}.%

\begin{figure}
[ptb]
\begin{center}
\includegraphics[
scale=0.39]%
{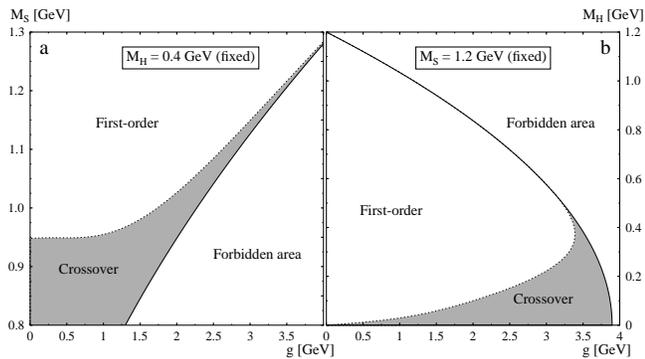}%
\caption{Order of the phase transition as function of the parameters of the
model. The forbidden area violates the constraint $\left\vert M_{S}^{2}%
-M_{H}^{2}\right\vert \geq4g\varphi_{0}$.
On the border line between the first-order and the crossover transitions a
second-order phase transition is realized.}%
\end{center}
\end{figure}

One observes in Fig.\ 1(a) that, in the limit $g\rightarrow0$ in which $S$ is
a pure quarkonium, the transition is crossover below a value $M_{S}%
\simeq0.948$ GeV and of first-order above it. The fact that a heavy
chiral partner of the pion induces a first-order chiral phase transition has
been discussed previously, see e.g.\ Ref.\ \cite{stefan}. This means
that, in a linear sigma model without tetraquark degrees of freedom, a heavy
(i.e., mass larger 1 GeV) chiral partner of the pion is excluded by lattice
QCD calculations \cite{karschfodor}, which indicate a crossover transition.
Including tetraquarks dramatically changes this conclusion: as shown in
Fig.\ 1(a), for increasing $g$ the region of crossover transitions extends
towards larger values of $M_{S}$. The scenario outlined in the
Introduction, in which $H$ $\equiv f_{0}(600)$ and $S\equiv f_{0}(1370)$, can
accommodate for a crossover transition if $g$, and thus the mixing between
quarkonium and tetraquark, is sufficiently large. Note, however, that the
range of $g$ values for which the transition is crossover, narrows
substantially when $M_{S}$ increases. We also find that, along the line of
second-order phase transitions in Fig.\ 1 (a), $T_{c}$ sizeably decreases for
increasing $g$, for instance from $T_{c}\simeq241$ MeV at $g=0$ to
$T_{c}\simeq186$ MeV at $g=3$ GeV and $T_{c}\simeq173$ MeV at $g=4$ GeV.
We observe in Fig.\ 1(b) that a crossover transition occurs only for small
values of $M_{H}$. The crossover region widens when $g$ increases. In order
to accommodate a value $\sim0.4$ GeV \cite{caprini}, a large value of $g$ is
required. Along the line of second-order transitions, $T_{c}$ first decreases,
and then increases for increasing $g$. The minimum $T_{c} \simeq145$ MeV
occurs for $g\simeq2$ GeV.

We now study the $T$-dependence of masses, condensates \cite{con}, and the
mixing angle in more detail in the case of $M_{H}=0.4$ GeV and $M_{S}=1.2$ GeV
(in the range quoted by Refs. \cite{pdg,pelaez,caprini}; a mass $M_H \sim 0.4$ GeV, altough leading to a too small tree-level decay width
 due to lack of phase space, allows for a nice illustrative
description of the qualitative features of the nonzero $T$). Also, we set the
coupling strength $g=3.4$ GeV, in order to obtain a crossover phase
transition in agreement with lattice QCD calculations \cite{karschfodor}.
These parameter values lead (together with
$\varphi_{0}=f_{\pi}=92.4$ MeV and $M_{\pi}=0.139$ GeV) to $M_{\chi}=0.82$
GeV, $M_{\varphi}=0.96$ GeV, and $\lambda=52.85$.%

\begin{figure}
[ptb]
\begin{center}
\includegraphics[
scale=0.39
]%
{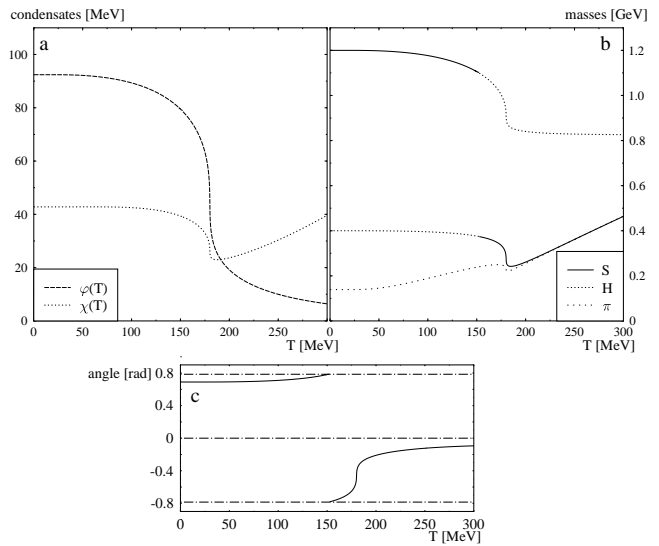}%
\caption{Condensates (panel a), masses (panel b) and mixing angle (panel c) as
function of $T.$}%
\label{fig2}%
\end{center}
\end{figure}

The condensates $\varphi(T)$ and $\chi(T)$ are shown in Fig.\ 2(a). The quark
condensate $\varphi(T)$ drops at $T_{c}$ and then approaches zero, signaling
the restoration of chiral symmetry. The tetraquark condensate $\chi(T)$ first
drops together with $\varphi(T)$, but increases above $T_{c}$: this is due to
the fact that in the equation determining $\chi(T)$, the growth of the $S$ and
$H$ tadpole contributions with $T$ has to be balanced by an increase of
$\chi(T)$; this could be different if we include additional terms $\sim
\chi^{4}$ in the original potential (\ref{vpot}). In any case, the increase of
$\chi(T)$ affects the behavior of the masses or of other physical quantities
only slightly. Note that the field $\chi$ is a singlet under chiral
transformations in the $N_{f}=2$ case and therefore the nonzero value of the
condensate does not imply a breaking of chiral symmetry at high $T$.

The $T$-dependent masses $M_{H}(T),$ $M_{S}(T)$, $M_{\pi}(T)$ of the
particles $H,$ $S$, and $\pi$ are shown in Fig.\ 2(b). The solid line
corresponds to $M_{S}(T)$, the mass of the state which is predominantly
quarkonium ($\left\vert \theta(T)\right\vert <\pi/4$), and the dotted line to
$M_{H}(T)$, the mass of the state which is predominantly tetraquark. At
$T_{s}\simeq160$ MeV, both masses behave discontinuously and the states
interchange their roles: for $T<T_{s}$, the state $S$ is the heavier scalar
and $H$ the lighter one, and for $T>T_{s}$, the state $S$ is lighter than $H.$
Above $T_{s}$ the state $S$ becomes degenerate with the pions as in the sigma
model without tetraquark. Note that, before becoming degenerate with the pion,
the thermal mass of the lightest state (identified with $H$ for $T\leq$
$T_{s}$ and with $S$ above it) slightly decreases, see also Ref. \cite{thermal}.

In Fig.\ 2(c) the mixing angle $\theta(T)$ is plotted. At $T_{s}$, $\theta(T)$
is discontinuous, which leads to the discontinuity in the masses noted above:
it jumps suddenly from $\pi/4$ to -$\pi/4$, $\lim_{T\rightarrow T_{s}^{\pm}%
}\theta(T)=\mp\frac{\pi}{4}.$ At $T_{s}$ the mixing is maximal: the two
physical states $H$ and $S$ have the same amount $(50\%)$ of quarkonium and
tetraquark. Note that, at $T_{s}$, the
field $\varphi$ and $\chi$ are degenerate in mass, which explains the maximal
value of the mixing angle. As a last remark we note
that the relative magnitude of $T_{s}$ and $T_{c}$ ($T_{s}<T_{c}$ as in our
example or vice versa) depends on the choice of the parameters. When
increasing the mixing strength $g$ the temperature $T_{s}$ decreases faster
than $T_{c},$ thus realizing the ordering $T_{s}<T_{c}$ in which the jump
occurs at smaller temperatures than chiral symmetry restoration.

\emph{Conclusions-- }In this Letter we proposed a novel scenario for chiral
symmetry restoration at nonzero $T$, in which two scalar-isoscalar states, a
tetraquark and bare quarkonium field, are considered. The mixing of the latter
two generates two physical states which can be associated with the resonances
$f_{0}(600)$ and $f_{0}(1370).$ When the tetraquark mass is smaller than the
quarkonium mass, as supported by various spectroscopic studies in the vacuum,
the state $f_{0}(600)$ is predominantly tetraquark
and $f_{0}(1370)$ is predominantly quarkonium. This scenario has
been studied by employing a simple model which includes only these two scalar
resonances and the pion triplet.

A remarkable aspect of our results is that the tetraquark-quarkonium mixing
generates a softer first-order phase transition or, depending on the coupling
strength, even a crossover transition. While in the standard linear sigma
model ($g=0$) a heavy chiral partner of the pion (with mass exceeding 1 GeV)
always leads to a first-order phase transition, this is not necessarily the
case when tetraquark-quarkonium mixing is considered: for sufficiently large
coupling strength $g$, the chiral transition is crossover, just like in
lattice QCD studies \cite{karschfodor}.
We also demonstrated that the mixing between quarkonium and tetraquark states
increases with increasing temperature, and, in most cases, reaches its maximal
value of $45^{\circ}$ at a temperature $T_{s}$ where the physical states
consist of an equal amount of quarkonium and tetraquark. For $T>T_{s}$ the
physical states interchange their roles: the lighter state is predominantly
quarkonium and the heavier predominantly tetraquark. Further increasing $T$
leads to the standard scenario of chiral symmetry restoration, where the
scalar quarkonium becomes degenerate in mass with its chiral partner, the
pion. Thus, our approach can possibly solve an inconsistency between
low-energy spectroscopy, where a non-quarkonium structure for $f_{0}(600)$ is
favoured, and studies at nonzero temperature, where the scalar partner of the
pion should be sufficiently light ($\sim0.6$ GeV) in order for the chiral
symmetry restoring transition to be crossover.

Since the present work is a first explorative study on the relevance of the
tetraquark at nonzero $T,$ we omitted other scalar-isoscalar states such as
$f_{0}(980),$ $f_{0}(1500)$, and $f_{0}(1710)$, which would naturally appear
in an $SU(3)$-symmetric model with two scalar nonets and a glueball state.
Also (axial-)vector mesons should be considered \cite{kr,stefan}. All
these fields are important in a more realistic framework which aims to
describe at the same time vacuum phenomenology and nonzero temperature
properties. We regard the results of this paper as a motivation to undertake
this more ambitious step in the near future.


\begin{thebibliography}{99}
\bibitem {con}The vev's $\varphi(T)$ and $\chi(T)$ can be
identified with the microscopic quark and 
tetraquark condensates by multiplying suitable powers of 
$\Lambda_{\text{QCD}}\sim250$ MeV: $\langle\overline{q}%
q\rangle\sim\Lambda_{\text{QCD}}^{2}\varphi(T)$ and $\langle
[\overline{q},\overline
{q}][q,q]\rangle\sim\Lambda_{\text{QCD}}^{5}\chi(T),$ see for instance Ref.
\cite{fariborz1}.
\end{thebibliography}
\end{document}